\documentstyle[epsf,aps,prl]{revtex}

\begin{document}

\twocolumn[\hsize\textwidth\columnwidth\hsize\csname
@twocolumnfalse\endcsname

\title{Multiwalled carbon nanotube: Luttinger liquid or not?}
\author{R. Tarkiainen$^1$, M. Ahlskog$^1$, J. Penttil\"a$^1$, L.
Roschier$^1$, P. Hakonen$^1$, M.
Paalanen$^1$,\\ and E. Sonin$^{1,2}$}
\address{ $^1$Low Temperature Laboratory, Helsinki University of
Technology,
FIN-02015 HUT, Finland \\
$^2$The Racah Institute of Physics, The Hebrew University of Jerusalem,
Jerusalem 91904, Israel}

\date{\today}
\maketitle

\begin{abstract}

We have measured $IV$ curves of multiwalled carbon  nanotubes using end
contacts.  At low voltages, the tunneling conductance obeys non-Ohmic power
law, which is predicted both by the Luttinger liquid  and the
environment-quantum-fluctuation theories. However, at higher voltages we
observe a crossover to Ohm's law with a Coulomb-blockade offset, which
agrees with the environment-quantum-fluctuation theory, but  cannot be
explained by the Luttinger-liquid theory. From the high-voltage tunneling
conductance we determine the transmission line parameters of the nanotubes.

\end{abstract}


%\draft

\bigskip

] %End of title in one column mode

Metallic carbon nanotubes are considered as outstanding realizations of
strongly
interacting, one-dimensional electron systems, {\it i.e.}   Luttinger liquids
(LL) \cite{General,Dekker,LL}.
A Luttinger liquid is a paramagnetic metal without  Fermi-liquid
quasiparticles.
Its basic charged excitations  are plasmons which can be viewed as
propagating
electrodynamic  modes in a similar fashion as in any regular
transmission
line.
Experimental evidence for LL behavior has recently been observed in
single
walled
carbon nanotubes \cite{Bockrath} as well as  in multiwalled tubes (MWNT)
\cite{Schonenberger99}. The transmission line analogy, in turn,
facilitates the
connection of LL theory to the environment-quantum-fluctuation (EQF) theory
\cite {Devoret,Ingold}.  This theory has been successful in explaining
Coulomb blockade
in normal tunnel junctions \cite{Ingold}. Unlike the LL
model, the EQF theory incorporates various factors, which makes it much more
amenable to
detailed experimental comparison, especially in the case of resistive
transmission
lines.

In this paper we present experimental results on the $IV$-curves for
four metallic, arc-discharge-grown  MWNTs.
We  analyze our results using both the EQF analysis
\cite{Devoret,Ingold}
as well as  the standard LL formulas \cite{Kane,Egger}. At  small
voltages both of
these approaches predict for $IV$-curves a  power law $I
\propto V
^{\alpha +1}$, which is also supported by the experiments yielding
$\alpha +
1=1.23 \pm 0.1$. At large
 voltages we find that only the EQF theory is applicable and obtain
for the
high frequency impedance $Z=1.3-7.7$ k$\Omega$. From the values of
$\alpha
$  and
$Z $ we can independently determine   the kinetic inductance of the
nanotubes
and  obtain consistent  values of $l_{kin}= 0.1-4.2$ nH/$\mu$m.

A multiwalled nanotube consists of several concentric nanotubes. About one
third of the tubes are expected to be metallic with quite large inter-layer
capacitance. According to our analysis, the large capacitance connects the
inductive as well the resistive  components of separate tubes in parallel. All
metallic tubes take part in the conduction at high frequencies, in contrast to
the Aharonov-Bohm experiments of Ref. \cite{Schonenberger99} where only the
outermost layer  contributed to the dc-resistance. The total electron density
$n$ of a MWNT is proportional to $M$, the number  of metallic layers, and  in
each channel the Fermi velocity of 1D  electron gas is $v_{F} = \pi \hbar
n/4Mm^* = 8\cdot 10^5 $m/s. Here $m^*$ is the effective mass
of an  electron, and we have taken into account the fact that
each metallic layer has four independent 1D conduction channels.

A metallic nanotube, placed on a silicon substrate between metallic
contact lines, can be viewed as an inner conductor of a transmission line
whose outer conductor is  formed by nearby metallic bodies. The
capacitance  per unit length of the line is $c =  2\pi \epsilon \epsilon_0
/$ln$(r_g/r_0)$. Here $\epsilon$ is the dielectric constant of the medium
between the conductors, $r_0$ is the outer radius of the nanotube, and $r_g$
is a distance from a metallic ground. The current carriers of the nanotube
occupy 1D conduction bands  and, in contrast to the carriers in metallic
wires, they have a low total density $n$, resulting in a large kinetic
energy stored in the current flow. Therefore, the magnetic inductance $l_m =
\mu_0 $ln$(r_g/r_0) /2\pi$, which is usually relevant for transmission
lines, has to be replaced by the kinetic inductance $l_{kin} =  m^*/ne^2$,
since $ l_{kin} \gg l_m$. In addition, for a 1D plasmon in
a nanotube  the inverse
compressibility $d\mu /dn = m^*{v_F}^2/n$ of the neutral Fermi gas
becomes comparable to the
electrostatic inverse compressibility  $e^2 /c$ of the transmission line
geometry. This can be taken into account by renormalization of the
nanotube capacitance into $\tilde c$:
\begin{equation}
{1\over \tilde c} ={1\over c}+ {1\over e^2 }{d\mu \over dn} = 
{1\over c} + {v_F^2 l_{kin}}~.
\end{equation}
Hence, the plasmon velocity $v_{pl}$ is
\begin{equation}
v_{pl}={1\over \sqrt{l_{kin} \tilde c}}=\left[1/l_{kin}c + v_{F}^{2}\right]
^{1/2}~,
\label{vpl}
\end{equation}
and in the expression for the line impedance $Z=V/I$,
\begin{equation}
 Z=\sqrt{l_{kin}/\tilde
c}=l_{kin}v_{pl} =\left[l_{kin}/c + (R_K/8M)^2\right] ^{1/2}~,
\label{LC}
\end{equation}
$V$ is the electrochemical (not only electric) 
potential difference, and $R_K=h/e^2$ is the quantum resistance.

In the above classical electrodynamic analysis the 1D plasmon modes are
the only excitations of the nanotube transmission line. The LL model for an
infinitely long  MWNT \cite{Kane,Egger} recovers the electrodynamic plasmon
mode with $v_{pl}$ given by  Eq. (\ref{vpl}) (cf. the  expression after Eq. (4)
in Ref. \cite{Kane}). But in addition to  the plasmon mode, the
Luttinger liquid has charge-neutral modes (a spin wave among them), which
propagate with the velocity different from  $v_{pl}$ and  keep the total charge
density constant. The Coulomb interaction, measured by the difference of
$v_{pl}/v_F$ from unity, suppresses the single electron density of state (DOS)
$\rho(E)=dn/dE$ near the Fermi level. The DOS is given by the Fourier component
of the electron Green's function $\langle \hat \psi(x,t) \hat
\psi(x,0)^\dagger\rangle$ ($\hat \psi(x,t)$ is the electron operator) and is
probed by the $IV$ curve: $dI/dV\propto\rho(eV)$.  At low energies
$\rho(E) \propto E^{\alpha_L}$, where  for  an end-contacted infinitely long
MWNT \cite{Egger,Matveev}
\begin{equation}
\alpha_L =( v_{pl}/v_F - 1)/4M ~.  \label{g}
     \end{equation}
In the limit of large $M$ or no-interaction $v_{pl}/v_F=1$, the
$IV$-characteristics of a MWNT approach Ohm's law ($\alpha_L=0$), $\it
{i.e.}$ the Luttinger liquid turns into a Fermi-liquid.

Another approach, the EQF theory, considers the effects of environment quantum
fluctuations on  $IV$-characteristics under the conditions of
Coulomb blockade. A non-resistive, infinitely  long nanotube acts as a
dissipative environment, {\it i.e.}, as a heat bath with which  the tunneling
electron can exchange energy \cite{Devoret,Ingold}. The energy exchange is
characterized by the function $P(E)$, which is a Fourier component of the
correlator $\langle e^{i\hat\varphi(t)} e^{-i\hat \varphi(0)}\rangle$, where
$\hat\varphi(t)$ is the operator of the phase.  At $T=0$, $P(E)$ is
proportional to the {\em second} derivative of the current: $d^2I/dV^2 \propto
P(eV)$.  In the Coulomb blockade regime, {\em i.e.} when the voltage bias  is
less than
$e/C_T$, where   $C_T$ is the capacitance of the  tunnel contact, the EQF
theory predicts that $P(E) \propto E^{\alpha_E -1}$ and $I\propto V^{\alpha_E
+1}$ with  $\alpha_E= 2 \mbox{Re}\{Z\}/R_K$. Using the impedance
$Z=\sqrt{l_{kin}/\tilde c}$ of the nanotube,  this yields the same power law as
the LL theory in the large $M$ limit. This fact, pointed out in Ref. \cite
{Matveev}, is not accidental. In the LL picture the current is suppressed
because there are no single electron quasiparticles, and the charge is
transported by bosonic modes (plasmons). Although in a junction between 3D
wires there  are single-electron states available (in contrast to 1D),  a
tunneling electron at $V << e/C_T$ has not enough energy to get into them. As
a result, the charge is transported again with 1D plasmons, which have similar
properties for 1D and thin 3D wires.

On the other hand, one should expect a similarity between $\rho(E)$ and
$P(E)$, since both the operators, $\hat \psi^\dagger $ and $e^{-i\hat\varphi}$,
which define these two functions, are creation operators for the
charge $e$. But if the exponents $\alpha_L$ and $\alpha_E$ for the conductance
coincide, the exponents $\alpha_L$ and $\alpha_E -1$ for $\rho(E)$ and
$P(E)$, respectively, differ by one. One can show \cite{S}, however,
that similar 
relations connect the $\rho(E)$- and $P(E)$-exponents  with the
impedance: $\alpha_L= 2 \mbox{Re}\{Z_L \}/R_K-1$ and $\alpha_E-1 = 2
\mbox{Re}\{Z\}/R_K-1$. But due to charge-neutral modes, which were not
considered in the EQF theory, the nanotube impedance differs in the LL theory
from the impedance $Z$ given by Eq. (\ref{LC}): $Z_L= Z+(4M-1)R_K/8M$. The
difference in the impedance compensates the difference in the relations
connecting  $\rho(E)$ and $P(E)$ with the conductance, and eventually in the
large $M$ limit both the theories predict the same exponent.

But the two approaches differ in their predictions for high voltages.
According to the EQF theory the power law is only valid in the Coulomb blockade
regime $\omega C_T << 1/Z $. The relevant frequency
$\omega =eV/\hbar$ in this inequality  corresponds to the environment mode
excited by a tunneling  event; in our experiments this means frequencies
up to
about 20 THz. At high frequencies and voltages  the  environmental
impedance $Z$
is shunted by tunnel junction capacitance $C_T$ and becomes $(i\omega C_T
+1/Z)^{-1}$. Then the EQF theory gives the  formula \cite{Ingold}
\begin{equation}
I={\frac{1}{R_{T}}}\left[ V-{\frac{e}{2C_{T}}}+\frac{R_{K}}{ Z} \left(
\frac{e}{2\pi C_{T}}\right) ^{2}{\frac{1 }{V}}\right]~.
 \label{IV-g} \end{equation}
This high-voltage asymptotics, characterized by the Coulomb offset $e/2C_T$ and
the ``tail'' voltage $\propto 1/V$  was experimentally studied and
discussed  by Wahlgren {\it
et al.}, and Penttil\"a {\it et al.}  within the horizon picture
\cite{WDH,Penttila00}. In contrast to the  EQF, in the LL  approach
the capacitance $C_T$ of the tunneling contact is absent,  and therefore
this approach does not predict a crossover to the  ``tail'' asymptotics given
by Eq. (\ref{IV-g}).

A summary of our four nanotube samples, each with a diameter of about 15
nm,
is presented in Table I. For contact,
we employed gold electrodes which were evaporated either prior to or
after the
deposition of nanotubes. Deposition of nanotubes was done as described
in
Ref. \cite{Leif99}. Mapping of nanotubes with respect to alignment marks
as
well as AFM micromanipulation was performed using Park Scientific
Instruments Autoprobe CP. Chrome or titanium (2-3 nm layer) was employed
as
an attachment layer before evaporating gold. Vacuum brazing at 700 C for
30
sec was employed to lower the contact resistance in samples T1-T3. On
the
dilution refrigerator,
the samples were mounted inside a tight copper enclosure and the
measurement
leads were filtered using 0.5 m of Thermocoax cable.

Tunnel junction capacitances $C_{T}=31-111$ aF and
resistances $R_{T}=20-68$ k$\Omega $ (neglecting the tube
resistance) were determined from
asymptotic behavior by fitting Eq. (4) to the
measured $IV$-curves. Owing to their relatively large size,
the contacts to the nanotube are not ideal
and may cause a small uncertainty in the interpretation
of the $\alpha$ values. Namely, the EQF theory and the LL model in the 
strong interaction and $M >> 1$
limit  predict two times smaller $\alpha$ values
for the bulk than for the end contact \cite {Ingold,Kane,Egger}.

Fig. 1 illustrates the low-voltage $IV$-curve of all four samples T1-T4.
We are plotting
the quantity $V_{offset}=V-I/\frac{dI}{dV}$ {\it vs.} $V$, which in
the case of a power law yields a straight line with the
slope $\alpha/(1+\alpha)$.  Only slight deviation of linear behavior
is seen at low voltages in Fig. 1. This indicates
that the Coulomb blockade of the island is rather
weakly seen (except in T4). The linear behavior also implies that the
two tunnel junctions become independent.
At voltages $\vert V \vert > 5$ mV, in spite of the additional wiggles,
slight tendency toward
saturation is observed in the data.
This is consistent with the EQF picture,
which predicts that
at high voltages $V_{offset}$ must gradually approach to a
$C_{T}$-dependent constant. By fitting a straight line
through each data set at $4 < \vert V \vert < 7$ mV, we obtain
 $\alpha =0.23\pm 0.1$ (Table 1).
Using Eqs. (2) and (3) we obtain 
$l_{kin_{\alpha}}=0.2-1.7$ nH/$\mu$m for the kinetic inductance.
Capacitance for the nanotube $\sim 70$ aF/$\mu$m is estimated using
the average of $c=2C_T /{\cal L}$ where we employ the total tube
length ${\cal L}$ for scaling.

Fig. 2 displays $IV$-curves measured at large voltages. In order to
facilitate a
direct comparison with the power-law dependence,
we have plotted our results on a log-log scale. Our
data are rather close to a single power law with small $\alpha$ but, at
larger values of $\alpha$ (samples T1 and T3),
there is a gradual approach
toward a linear law as expected for a single junction in a resistive
environment. Thus both figures give evidence that the environmental
(EQF) theory is better suited for the analysis at high voltages. In fact,
also the saturation observed by Bockrath {\it et. al.} \cite{Bockrath} can
be explained by asymptotic approach toward Ohm's law.

The plasma resonances, which one expects in finite nanotubes,
are washed away in our samples. This gives a lower limit for the resistivity of
the line, $r{\cal L}\geq Z/4$.  On the other hand, the $LC$-line model works
over our voltage range, {\it i.e.} $ r\leq \omega l_{kin}$, which results in an
upper limit for $r$ of the order of 1  k$\Omega / \mu$m (at 1 mV). For
comparison, from the two-terminal resistance measurements we
estimate that $r \lesssim 20$ k$\Omega /\mu $m for our tubes at DC \cite{note}.

One may argue that the poor agreement of the LL picture with the high-voltage
part of the $IV$ curves could be reconciliated by 
including the junction capacitance
$C_T$ into the impedance, as is done in the 
EQF theory. There is, however, a conceptual
problem to do it. The density of state $\rho$ is expected to be a bulk
property and, therefore, independent of $C_T$. Moreover, inclusion of $C_T$
into the impedance, which determines $\rho$, does not help to match the
LL picture with experimental results. 
The capacitance $C_T$ short circuits the environment
impedance, and  $\rho(E)$ should decrease with $E$,
like $P(E)$ in the EQF theory.  But since $dI/dV \propto \rho(E)$, in contrast
the EQF theory where $dI^2/dV^2 \propto P(E)$, this yields  a high-voltage
plateau (voltage-independent current), but not Ohm's law with Coulomb offset.
Introduction of a proper high-energy cut-off in the LL model could explain a
cross-over to Ohm's law, but
not a Coulomb offset. We expect this cut-off to be larger than the
region of our analysis which is bounded by the presence of higher transverse
modes above 50 mV.

Fits, based on Eq. (\ref{IV-g}), fall on top of the experimental
data in Fig. 2. In the fitting, we assume that the junctions at the ends of the
tube are symmetric and, in fact,
$I$ $vs.$
$V/2$ is fitted to the single junction formula. We
also tried to incorporate a cubic background, $
\eta V^{3}$ in the fitting, which was found essential in Al-samples because
of the deformation of the tunnel barrier at high voltages \cite{Penttila00}.
Surprisingly, the cubic
term was found negligible in all our nanotube samples. Our
fits yield a characteristic impedance of $Z=1.3-7.7$
k$\Omega$ for the resistive environment. These results depend slightly
on the
measurement polarity (see Table I). Finally, using Eqs. (2) and (3), we
obtain for the kinetic inductance $l_{kin}=0.1-4.2$ nH/$\mu$m.

Table I contains parameters obtained both
from the EQF analysis for a $LC$-transmission
line as well as from the power law exponents according to the
LL model. The results of the two methods overlap each
other; the scatter of the power law analysis is
slightly smaller than that of the environmental analysis. 
In addition, we checked that the temperature dependence of the measured conductance,
$dI/dV \propto T^{\alpha} $, yielded consistent values of $\alpha = 0.25 \pm 0.1$.
As a final
result of all our determinations we quote the median
value $l_{kin}=0.5$ nH/$\mu$m. If we compare this with
the theoretical prediction $l_{kin} = R_K /8M
v_F$, we conclude that the average number of conducting
layers in our nanotubes is 8 and the large variation
of the inductance may come from the variation in $M$. The
average value of 8 indicates that about every
3rd layer in our nanotubes is metallic.

To conclude, on the basis of our experimental results we argue
that, at high voltages, the environmental theory gives a better account
of transport measurements of multiwalled nanotubes than the Luttinger
liquid picture, because the tunnel junction
capacitance is neglected in the Luttinger liquid theory. At lower voltages, no
distinction between these two theories can be made.  Due to
their large kinetic inductance, nanotubes provide an excellent high-impedance
environment for normal junctions at high frequencies, which
is crucial for single-electronics phenomena. As the kinetic inductances of
different layers are in parallel in MWNT, these phenomena will be more
pronounced in single walled carbon nanotubes.

We thank C. Journet and P. Bernier for supplying us with
arc-discharge-grown nanotubes. Interesting discussions with B. Altshuler, F. Hekking,
G.-L. Ingold, A. Odintsov, and A. Zaikin are gratefully acknowledged.
This work was supported by the Academy of Finland, by the Israel Academy of
Sciences and Humanities, and by the Large Scale Installation Program ULTI-3 of
the European Union.

%\newpage

\begin{table}
\caption{Summary of our samples of single, metallic tubes T1-T4.
  ${\cal L}_1$/${\cal L}_2$ denotes the length of the tube over
the metallic leads/the length of the free
standing section that is hanging
  $\sim 20$ nm above the substrate. For sample T4, ${\cal L}_2$
  specifies the length in contact with SiO$_{2}$. In the resistivity
ratio $rr =R_{0}$(4.2K)/$R_{0}$(290K), the values of $R_{0}$ have been
obtained from slopes of $IV$-curves at $I=0$. $T$ denotes the
measurement temperature for the data in Figs. 1 and 2. The junction
capacitance
$C_{T}$ and tunneling resistance $R_{T}$ are taken from
the $IV$-curve fits using Eq. (5).
These fits also yield the
lumped-element environmental
impedance $Z$ which is slightly different for
positive and negative voltages. Kinetic
inductance obtained from $Z$ using $c=70$ aF/$\mu$m and Eq. (3)
is given by $l_{kin_Z}$. The Luttinger liquid power
$\alpha$ is determined at $4 < \vert V \vert < 7$ mV.
The kinetic inductance $l_{kin_{\alpha}}$ is
an estimate obtained using Eq. (3).
Voltage range of the tail fits is given in the last column. }
\bigskip

\begin{tabular}{cccccccccccc}
 & ${\cal L}_1$/${\cal L}_2$ & $ rr $ & $T$ & $R_{T}$ &
$C_T$ & $Z$ & $l_{kin_Z}$ & $\alpha $ & $l_{kin_{\alpha}}$ &  Range
&  \\
& $\mu $m &  & K & k$\Omega $ & aF & k$\Omega $ & mH/m &  &
mH/m
& mV &  \\
T1  & 0.7/0.5 & 1.7 & 4.2 & 25 & 31 & 3.5/7.7 & 0.9/4.2 & 0.30 & 1.1 &
10-50 &
\\
T2  & 0.3/0.8 & 3.0 & 0.1 & 20 & 33 & 1.3/2.3 & 0.1/0.4 & 0.12 & 0.2 &
15-50
& \\
T3  & 0.8/0.6 & 4.6 & 4.2 & 46 & 37 & 2.0/4.8 & 0.3/1.6 & 0.32 & 1.2 &
10-50 &
\\
T4  & 1.4/2.3 & 3.0 & 0.1 & 68 & 111 & 1.8/2.7 & 0.2/0.5 & 0.17 & 0.3 &
7-20 &
\end{tabular}

\end{table}

\begin{figure}

\begin{center}
\epsfxsize=75mm
\epsffile[80 200 470 600]{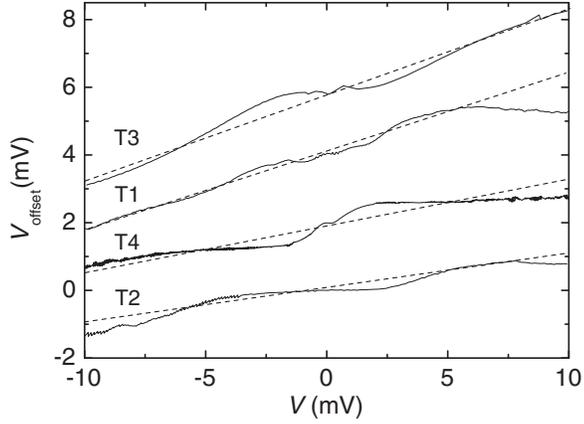}
\caption{Offset voltage $V_{offset}=V-I/\frac{dI}{dV}$ {\it vs.} $V$
for all our samples T1-T4; the power law behavior $I \propto V
^{\alpha +1}$ yields a straight line in
this kind of a plot. The dashed lines illustrate linear fits made in the
range
 $4 < \vert V \vert < 7$ mV. The effect of Coulomb
blockade near zero is seen to be
small except for the sample T4.} \label{f1}
\end{center}
\end{figure}

\begin{figure}
\begin{center}
\epsfxsize=75mm
\epsffile[80 200 470 600]{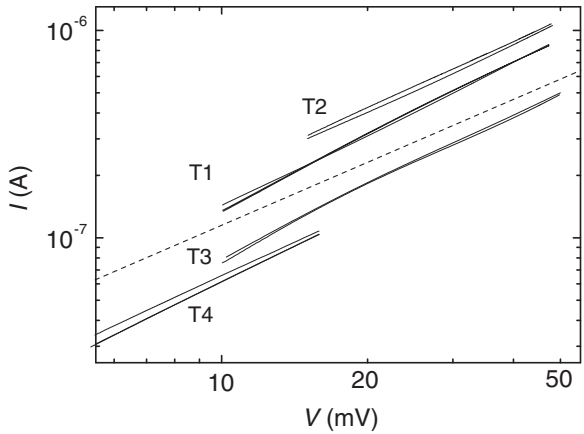}
\caption{High voltage $IV$-curves (both positive and negative polarities)
on a log-log plot. The dashed line
  illustrates linear behavior ($\alpha=0$). For details, see text.}
\label{f2}
\end{center}
\end{figure}


\begin{references}
\bibitem{General}  See, {\it e.g.}, ''Special Issue on Nanotubes'' in
Physics World, June 2000, p. 29.

\bibitem{Dekker}  C. Dekker, Physics Today, May 1999, p. 22.

\bibitem{LL}  See, {\it e.g.}, M.P.A. Fischer and L.I. Glazman
in {\it Mesoscopic Electron Transport},
  edited by L.L. Sohn, L.P. Kouwenhoven, and G. Sch\"{o}n (Kluwer
  Academic Publishers, Dordrecht, 1997) p. 331; J. Voigt,
cond-mat/0005114.

\bibitem{Bockrath}  M. Bockrath, D.H. Cobden, J. Lu, A.G. Rinzler, R.E.
Smalley, L. Balents, and P.L. McEuen, Nature {\bf 397}, 598 (1999).

\bibitem{Schonenberger99}  C. Sch\"{o}nenberger, A. Bachtold, C. Strunk,
J.-P. Salvetat, and L. Forro, Applied Physics A {\bf 69}, 283 (1999);
A. Bachtold, C. Strunk, J.-P. Salvetat, J.-M. Bonard,
L. Forro, T. Nussbaumer, and C. Sch\"{o}nenberger, Nature {\bf 397},
673 (1999).

\bibitem{Devoret}  M.H. Devoret, D. Esteve, H. Grabert, G.-L. Ingold, H.
Pothier, and C. Urbina, Phys. Rev. Lett. {\bf 64}, 1824 (1990).

\bibitem{Ingold}  G.-L. Ingold and Yu.V. Nazarov, in: {\sl Single Charge
Tunneling}, ed. H. Grabert and M.H. Devoret, (Plenum Press, N.Y., 1992),
pp.
21-107.

\bibitem{Kane}  C. Kane, L. Balents, and M.P.A. Fischer,
  Phys. Rev. Lett. {\bf 79}, 5086 (1997).

\bibitem{Egger}  R. Egger, Phys. Rev. Lett. {\bf 83}, 5547 (1999).

\bibitem{Matveev} K.A.  Matveev and L.I. Glazman,
  Phys. Rev. Lett. {\bf 70}, 990 (1993).

\bibitem{S} E.B. Sonin, cond-mat/0103017.

\bibitem{WDH}  P. Wahlgren, P. Delsing, and D.B. Haviland,
  Phys. Rev. B
{\bf 52}, R2293 (1995); P. Wahlgren, P. Delsing,
T. Claeson, and D.B. Haviland,
Phys. Rev. B {\bf 57}, 2375 (1998).

\bibitem{Penttila00}  J.S. Penttil\"{a}, \"{U}. Parts, P.J. Hakonen,
M.A.
Paalanen, and E.B. Sonin, Phys. Rev. B {\bf 61}, 10890 (2000).

\bibitem{Leif99}  L. Roschier, J. Penttil\"{a}, M. Martin, P. Hakonen,
M. Paalanen, U. Tapper, E. Kauppinen,
C. Journet, P. Bernier, Appl. Phys. Lett. {\bf 75}, 728 (1999).

\bibitem{note} Ballisticity of our tubes is enhanced by the
  AFM-manipulation
which seems to clean the surface. This is in accordance with
the work of  S. Frank, P. Poncharal, Z.L. Wang, W. A. de Heer, Science
{\bf 280}, 1744 (1998) in which ballistic propagation in MWNTs
was observed after dipping in to Hg-bath.

%\bibitem{cap}  We have
%estimated the inter-layer capacitance using the standard coaxial cable
%formula $c_{int}=\frac{2\pi \varepsilon ^{\prime }\varepsilon_0 }{\ln
%(b/a)}$ for two
%conducting layers separated by 1 nm of dielectric material with
%$\varepsilon
%^{\prime }=3$.

\end{references}
\end{document}